# View Optimization in Computed Tomography

**Seemeen Karimi[1] and Christopher McGuire[1]**

[1]Lawrence Livermore National Laboratory
7000 East Avenue
Livermore, CA 94550
(617) 953-1662; karimi1@llnl.gov

## ABSTRACT

Many emerging CT imaging applications must operate with data from sparse projection angles, on the order of 10 views, such as first responder scenarios, dynamic imaging of detonation events, and flash imaging of materials under extreme conditions in high energy-density science. Few-view imaging is a data undersampling problem, that degrades tomographic reconstructions with streaks, shadows, and other aliasing artifacts, due to ill-conditioning. In few-view CT, the specific set of projection angles can strongly impact reconstruction quality, and the optimal angles depend on the object being scanned. Despite a surge of recent research, there is much to explore in this problem. We present an object-aware performance prediction framework for few-view projection datasets based on a matrix decomposition of the discretized forward model augmented with expected object gradient information. The method yields a quantitative score for a candidate set of view angles that can underpin a view selection recommendation algorithm. A key feature is that it can be applied at multiple stages of acquisition: a priori, before any radiographs are taken, or adaptively, after some radiographs are collected, to determine the next best view. We demonstrate the approach using simulated CT data with 5 view and 15 view configurations. Preliminary results are encouraging; images reconstructed using a standard statistical algorithm show that reconstruction accuracy agrees with our performance score. Reconstruction accuracy was evaluated against a known ground truth using an information theoretic score. Continued testing with complex objects and multiple geometries is necessary.

## INTRODUCTION

In conventional x-ray computed tomography (CT) systems, an object is scanned with a dense set of angular measurements. Industrial, medical, and security systems commonly acquire hundreds to thousands of radiographs at different angles, called views, and reconstruction algorithms can recover volumetric structure sufficiently accurately. However, emerging imaging applications must work with sparse data when an object to be studied cannot be placed in a conventional CT scanner. For example, dynamic studies that necessitate the use of flash radiography can only have a few views. Ultrabright light sources, such as the linac coherent light source (LCLS) at SLAC [1], the European x-ray free electron laser (EuXFEL), and Advanced Photon Source at Argonne National Laboratory enable multi-view instruments with up to five pairs of sources and detectors for imaging materials under extreme conditions [2]. Lawrence Livermore National Laboratory (LLNL) and the Air Force Research Laboratory have imaging instruments for high-explosives safety and performance, that include 15 pairs of sources and detectors on the surface of a sphere [3]. The firing sequence of these beamlines is changeable and part positioning is flexible. In emergency response situations, first responders must acquire sequentially radiographs using a source and detector that is manually or partly manually positioned [4]. Only a few radiographs can be obtained because of the urgency of the situation and the threat to the first responders themselves. Other applications include large-part inspection, low-dose biological imaging, and inspection of electronic components.

The sparsity of views is a data undersampling problem and leads to a loss of detail and artifacts in the reconstructed tomographic image, such as streaks, shadows and other recognizable aliasing artifacts. Statistical reconstruction algorithms have been developed to converge to the most desirable local minimum [5][6], but are eventually limited by the information content of the set of radiographs. In systems with such few-views, the set of view angles have a critical impact on the CT reconstruction. The optimal view angles also depend on the object being scanned.

Research in view-selection has been rapidly gaining momentum in recently. There are geometric methods that evaluate the richness of a set of views in terms of the completeness of scanning trajectories [7]. These methods do not utilize any known information about the object being scanned. Another shortcoming is that they cannot adequately distinguish between one sparse set of angles and another. One method minimizes a covariance loss given a known object to determine the complementarity of views, a heuristic that may not include theoretical completeness criteria [8]. Another paper proposed a greedy scanning algorithm that selects the next best view based on a golden section sampling. This approach does not evaluate the views altogether, so it may miss the global optimum set of views, and for a fixed set of views, does not provide the best object placement and uses a burdensome real-time reconstruction [9].

We propose a performance prediction method for a set of view angles that is designed for imaging problems with very few views, is based on invertibility, and that also depends on the object selected. The method is based on singular value decomposition (SVD) of a matrix representing a discretized forward projection model. Object dependence is included by computing the gradient of a surrogate object at each voxel and cross-multiplying with detector normals. This method serves as the basis for an adaptive view recommendation algorithm. An advantage of this method is that it can be used anytime in the acquisition of a sequential imaging system: before any radiographs are taken, or after some radiographs are taken, when the next best view must be determined. We demonstrate the usefulness of our predictive method with simulated CT data for 5-view and 15-view scanning geometries, showing that the accuracy of reconstruction corresponds well with the prediction. Our main contributions are a new approach to determine the collective effectiveness of a few-view CT geometry using the SVD and a novel modification of the forward operator based on object gradients. This work examines the problem in a new way and puts forth a new heuristic. While our experiments show that our method correlates with reconstruction quality, the results are preliminary, and this framework must be further explored, tested against different geometries and objects, benchmarked for performance and tested with real scanner data.

## METHODS

We propose an algebraic spectral analysis method to compute the usefulness of a set of radiographic views based on SVD of the projection operator. The object and detector are discretized into voxels and pixels respectively, therefore, the projection operator can be expressed as a matrix, whose rows represent detector pixels, and whose columns represent the voxels of the object volume. Each cell of the matrix represents the contribution of an image voxel to a projection ray. We express the forward model as $p = Af + n$, where $f$ represents voxels, $p$ represents image voxels and $n$ represents noise. In doing so, we linearize the problem, neglecting polychromaticity and scatter. The matrix $A$ is determined by the tomographic geometry, and determines the invertibility of the problem. The null space of the matrix corresponds to the incompleteness of projection data. $A$ is decomposed as

$$A = U \Sigma V^T,$$

where $U$ and $V$ contain orthonormal left and right singular vectors, and $\Sigma$ contains the singular values $\sigma_i$. The singular values indicate the strength with which modes in the object space are represented in measurement space. All singular matrices, that is, $A$ in all sparse view systems, will have singular values that go to zero, but the rate of decay of singular values is an indicator of the ill-conditioning because it signifies the additional information in additional measurements. While we do not use the SVD to invert $A$ and obtain $f$, faster decays in $\Sigma$ correspond to a poorer invertibility. Our predictive score is the sum of singular values, i.e., the trace of $\Sigma$.

When a surrogate object is known, or its expected gradients are known, we can incorporate the gradients into our score through the projection matrix as follows. We compute the 3-dimensional gradients of the surrogate objects, and calculate the cross-product of the gradient at each voxel with the detector planes, or with the normal to each ray. We then calculate a function of the cross product for each voxel over all detector orientations and multiply it by the projection matrix. Let the object gradient at voxel $j$ be

$$\bar{g}_j = \nabla f_j$$

$$\bar{k}_{(i,j)} = \bar{g}_j \times \bar{d}_i$$

$$\tilde{A}_{i,j} = A_{i,j}\varphi(j)$$

The magnitude of $\bar{k}_{(i,j)}$ increases with the strength of the object gradient as well as how well it was captured by that projection ray. A reasonable choice for the function φ is a maximum over all projection rays:

$$\varphi(j) = \max_i \left|\bar{k}_{i,j}\right|$$

Voxels with gradient of zero do not contribute object-edge information. The corresponding columns can be removed from $\tilde{A}$, reducing computational cost.

To validate the method, we simulated data for 5 view and 15 view problems using the LEAP simulation and reconstruction suite [10]. For the five-view problem, we simulate two sets of views, as shown in Figure 1. We refer to these as "xfel" and "orthogonal" geometries. The xfel geometry corresponds to the geometry of a real XFEL instrument. The red lines represent the 5 detector normals and the blue dot shows the origin of the scanner's coordinate system. For practical computation, the number of voxels can be kept small, for example by downsampling, as long as the large gradients that would cause artifacts are captured. In the five-view example, the original problem had approximately $500^3$ voxels and $500^2$ detector pixels per view, rendering $A$ too large to decompose, so we downsampled the voxel and detector dimensions by a factor of ten in each spatial dimension. The resulting size of $A$ was $12500 \times 125000$. When the object was included, removing columns corresponding to the zero-gradient voxels further reduced the columns to about 6000. For the fifteen-view problem, we compared geometry of our 3-ring flash scanner with a modified version of it. Here the size of the A matrix was $37500 \times 125000$ before dropping any columns. We used an NVIDIA H100 accelerator with 80 GB of HBM3 memory and the Python CuPy library for SVD calculations.

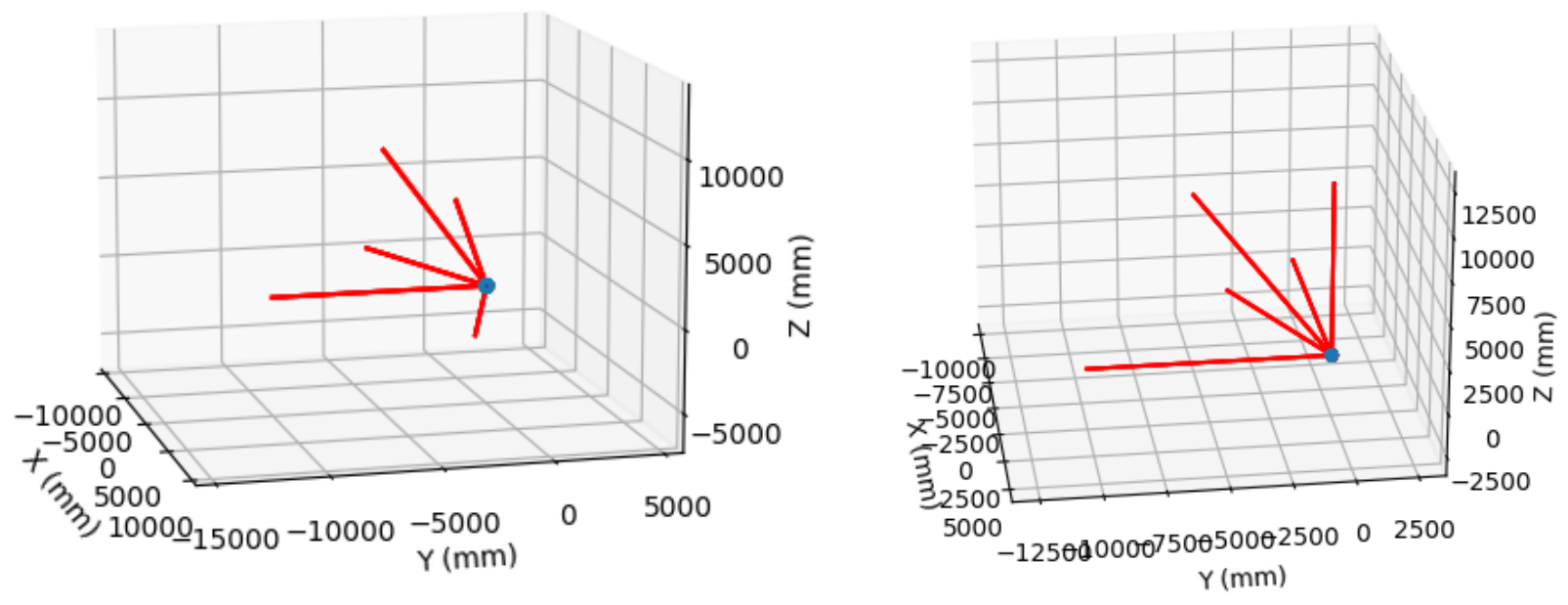

(a) (b)

**Figure 1. Two geometries: “xfel” (a) and “orthogonal” (b).**

## RESULTS AND DISCUSSION

We simulated forward projections of a parallelopiped with bubbles, with the two geometries shown in Figure 1. Images reconstructed with the well-known ASD-POCS algorithm [5] are shown in Figure 2. As we expect intuitively, the orthogonal geometry simulation resulted in better reconstruction than the xfel geometry. The edge contrast is better and the holes are reconstructed more accurately.

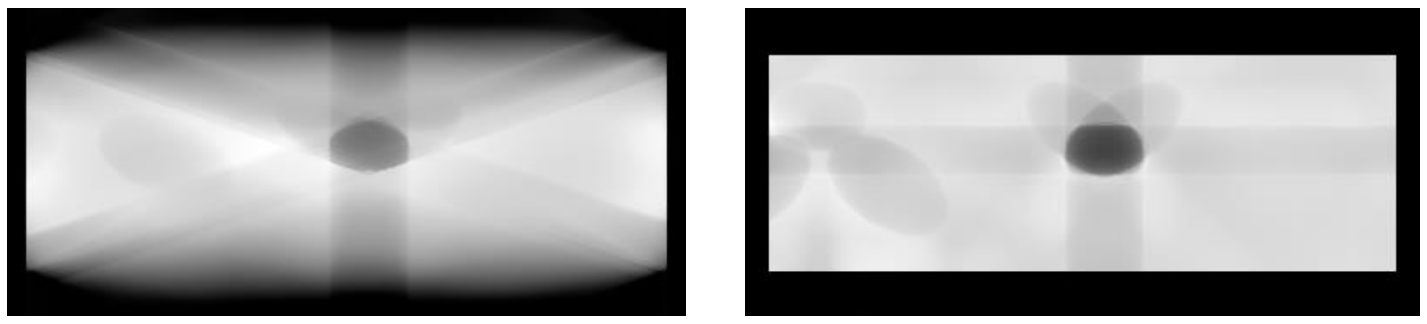

(a) (b)

**Figure 2. Reconstructed images from the xfel geometry (a) and the orthogonal geometry (b)**

Figure 3 shows the singular values from decomposing the projection matrices $A$ for the two geometries. The orthogonal geometry has the higher score than the xfel geometry. We use a surrogate object to calculate $\tilde{A}$, a solid parallelopiped of a different shape, but whose planes were at the same orientations as in the original parallelopiped. The disparity between the xfel and orthogonal geometries increases in the lower modes by including the object.

For the fifteen-view problem representing the 3-ring flash scanner at LLNL, we simulated data of the FORBILD head phantom and computed the gradients of a downsampled version of this phantom. During the downsampling, the smaller structures were lost. We moved two views from the 3-ring geometry so that the gradients along the x-direction (horizontal) and the perpendicular in the plane of the page were captured better. We simulated data using the modified geometry. Figure 4 shows images reconstructed from the simulated data from the original and modified geometries. The modified geometry yields better images, visually and numerically as shown in Table 1, where the multi-class F1 score and the L1 error improve by moving the views. Also shown in Figure 4 are the singular values of the decompositions of the object-modified matrices (the singular values of the original matrices are not shown). The difference between the sum of singular values increases from 3 to 4 percent when the projection matrix is modified by the phantom, so again, the discriminative power increases when using the object.

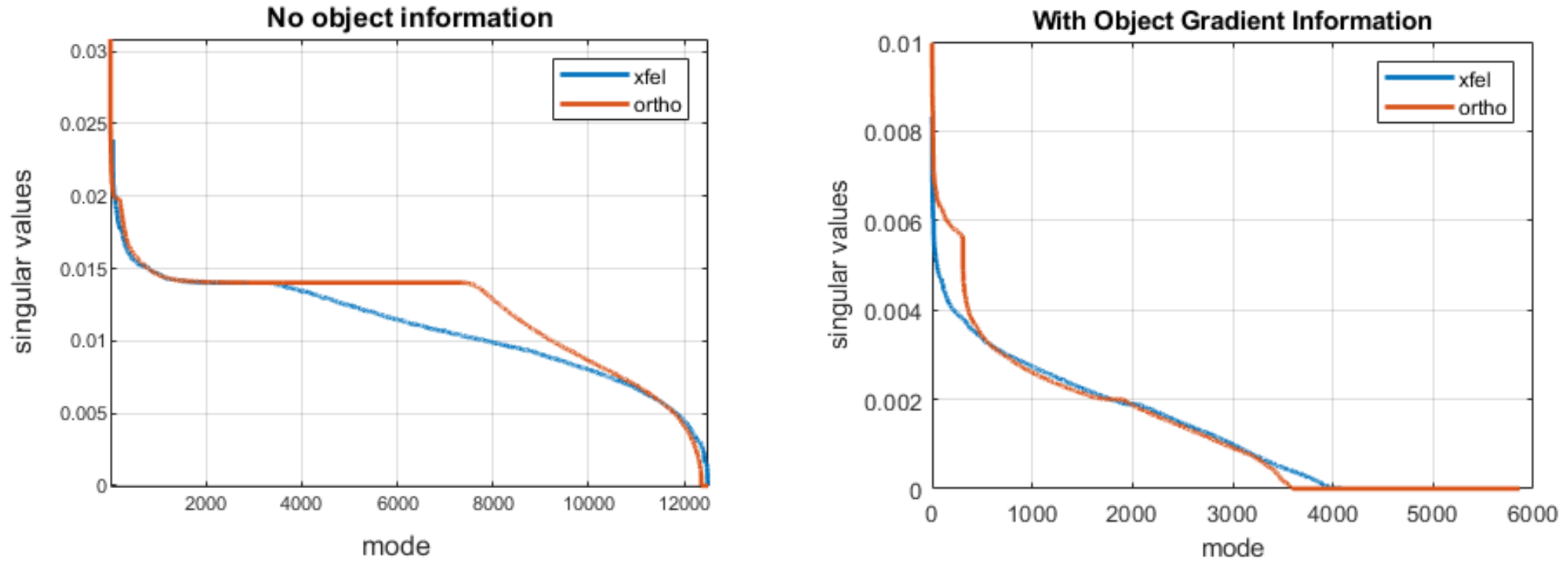

(a) (b)

**Figure 3. The singular values obtain from decomposing A (a) and $\widetilde{A}$ (b) for the two five view geometries**

The method works despite the downsampling factor of 10 in each dimension, because the larger edges that cause the undersampling artifacts are preserved. When the geometry that favors these larger edges is used for reconstruction, the artifacts do not obscure the smaller features as much. Therefore, the smaller features do not need to be preserved for decomposition.

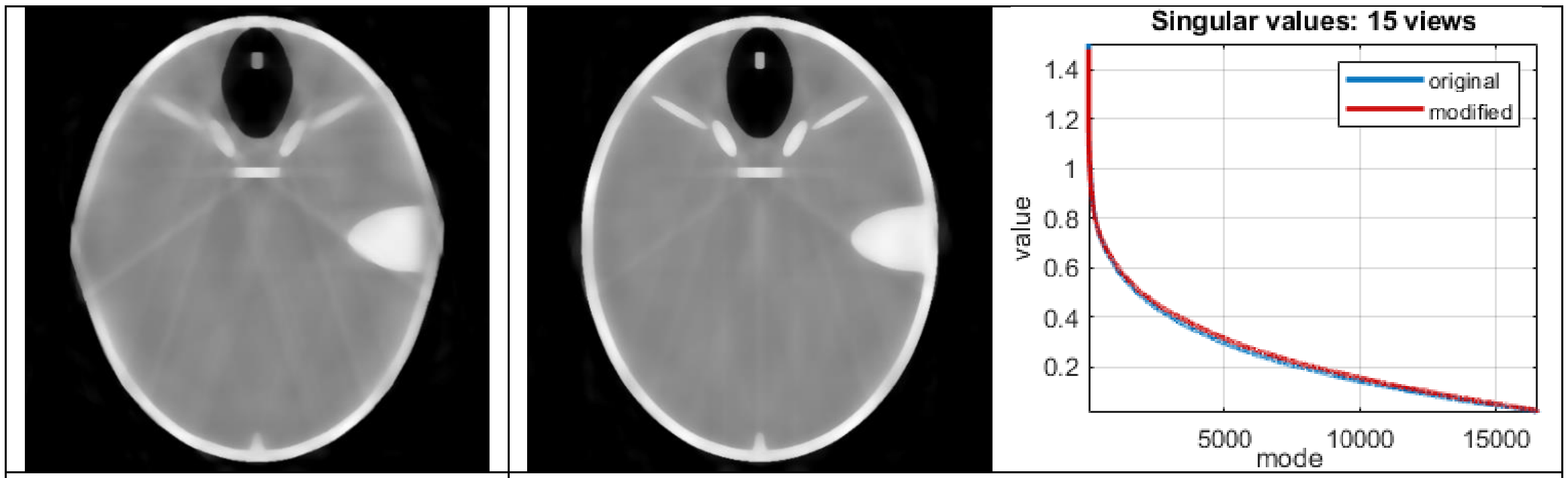

(a) (b)

**Figure 4. Reconstructions of the object from two different 15 view geometries. The better reconstruction (b), corresponds to the red curve of the plot, and greater singular values.**

**Table 1 Accuracy of the reconstructed image with respect to the ground truth for the head phantom.**

| **Measure** | **3-ring** | **Modified** |
|---|---|---|
| F1, multiclass | 0.268 | 0.284 |
| L1, normalized | 0.1 | 0.063 |

Many other aspects of this work need further study. It is necessary to test this approach on many different geometries, object shapes and features. We can use a non-linear function of $\bar{k}_{(i,j)}$ rather than $\bar{k}_{(i,j)}$ itself. We should investigate other functions instead of the sum of singular values. Instead of computing all singular values, we can find only the first K singular values where K must be researched. Our preliminary work has been on ideal (monochromatic, noise-free, and scatter-free) simulated data to isolate the view sampling problem from other issues. However, our method should be tested with real data.

## CONCLUSION

We have developed a novel predictive score for few-view geometries to compare their prospective image quality. This score incorporates knowledge about the object, so that it can be used to determine the best geometry for that object. It can be used as the basis of a recommender algorithm for adaptive few view selection and for optimizing object placement. We demonstrate this predictive score on a few geometries and a simple object. Additional validation with realistic physics, measured data, diverse geometries, and uncertainty-aware object models is needed before operational deployment. Nevertheless, the initial results indicate that object-dependent spectral analysis is a promising tool sparse-view CT systems.

**ACKNOWLEDGMENTS**

This work was performed under the auspices of the U.S. Department of Energy by Lawrence Livermore National Laboratory under Contract DE-AC52-07NA27344 and was supported by the Laboratory Directed Research and Development Program under LLNL-LDRD 24-ERD-021.